\documentclass{article}
\usepackage{array,booktabs}
\usepackage{graphicx}
\usepackage{comment}
\usepackage{longtable}
\usepackage{colortbl}
\usepackage{algorithm2e}
\usepackage{url}
\usepackage{amsthm}
\usepackage{tikz}
\usepackage{amsmath}
\usepackage{amssymb}
\usepackage{float}
\usepackage[caption=false]{subfig} 
\usepackage{enumerate} 
\usepackage[margin=1in]{geometry}

\usepackage
{siunitx} 

\newcolumntype{L}{@{}>{\kern\tabcolsep}l<{\kern\tabcolsep}}
\newcolumntype{R}{@{}>{\kern\tabcolsep}r<{\kern\tabcolsep}}
\newcolumntype{C}{@{}>{\kern\tabcolsep}c<{\kern\tabcolsep}}
\graphicspath{{./}}

\def\BibTeX{{\rm B\kern-.05em{\sc i\kern-.025em b}\kern-.08em
    T\kern-.1667em\lower.7ex\hbox{E}\kern-.125emX}}

\usetikzlibrary{decorations.pathreplacing,calc}

\newcommand{\tikzmark}[2][-3pt]{\tikz[remember picture, overlay, baseline=-0.5ex]\node[#1](#2){};}

\tikzset{brace/.style={decorate, decoration={brace}},
	brace mirrored/.style={decorate, decoration={brace,mirror}},
}

\newcounter{brace}
\newcommand{\drawbrace}[3][brace]{%
	\refstepcounter{brace}
	\tikz[remember picture, overlay]\draw[#1] (#2.center)--(#3.center)node[pos=0.5, name=brace-\thebrace]{};
}

\newcounter{arrow}
\newcommand{\annote}[3][]{%
	\tikz[remember picture, overlay]\node[#1] at (#2) {#3};
}
\theoremstyle{plain}
\newtheorem{mydef}{Definition}
\newtheorem{myth}{Theorem}
\newtheorem*{myth*}{Theorem}

\newcommand{\tss}[1]{\textsubscript{#1}}

\newcommand{\changed}[1]{{#1}}

\newcommand{\metaset}[1]{\ensuremath{\mathbb{#1}}}
\newcommand{\set}[1]{\MakeUppercase{#1}}
\newcommand{\leaveslist}{L}
\newcommand{\prefix}{p}
\newcommand{\rdxtreeset}{\metaset{T}}
\newcommand{\rdxtree}{T}

\newcommand{\pfxsetbgp}{P\tss{BGP}}
\newcommand{\opset}{\set{o}}
\newcommand{\opsets}{\metaset{O}}

\newcommand{\igpgraph}{G}
\newcommand{\route}{R}
\newcommand{\bgpnh}{n}
\newcommand{\link}{l}
\newcommand{\weight}{w}
\newcommand{\igpdists}{D}

\newcommand{\igpcost}{$\alpha$}
\newcommand{\precbgp}{$\prec~$}
\renewcommand{\P}{\ensuremath{\mathbb{P}}} 
\newcommand{\ie}{\emph{i.e.}} 
\newcommand{\eg}{\emph{e.g.}} 

\SetKw{hash}{hash}
\SetKw{add}{add}
\SetKw{fiprs}{MR\_protection\_index}
\SetKw{free}{free\_unused\_sets}
\SetKw{from}{from}
\SetKw{getnrs}{get\_MR}
\SetKw{ispro}{2disj\_routes}
\SetKw{kwbreak}{break}
\SetKw{kwcontinue}{continue}
\SetKw{kwith}{with}
\SetKw{kwin}{in}
\SetKw{kwor}{or}
\SetKw{maxrs}{max\_rs}
\SetKw{medsucc}{med\_succ}
\SetKw{kwmin}{min}
\SetKw{remove}{remove}
\SetKw{rep}{replace}
\SetKw{spf}{spt}
\SetKw{spfcmpnts}{spf\_elts}
\SetKw{upOP}{update\_OPR}
\SetKw{rg}{residual\_graph}
\SetKw{kwpath}{path}
\SetKw{kwEOS}{extractOPRSet}
\SetKw{kwismin}{is\_min}
\SetKw{kwusefulMR}{useful\_MR}

\newcommand\CR[1]{#1}

\makeatletter
\def\blfootnote{\xdef\@thefnmark{}\@footnotetext}
\makeatother

\begin{document}
%

\title{A Fast-Convergence Routing of the Hot-Potato}


\newcommand\copyrighttext{%
  \footnotesize \textcopyright 2021 IEEE INFOCOM. Personal use of this material is permitted.
  Permission from IEEE must be obtained for all other uses, in any current or future
  media, including reprinting/republishing this material for advertising or promotional
  purposes, creating new collective works, for resale or redistribution to servers or
  lists, or reuse of any copyrighted component of this work in other works.
  DOI: To Be published in INFOCOM}
\newcommand\copyrightnotice{%
\begin{tikzpicture}[remember picture,overlay]
\node[anchor=south,yshift=10pt] at (current page.south) {\fbox{\parbox{\dimexpr\textwidth-\fboxsep-\fboxrule\relax}{\copyrighttext}}};
\end{tikzpicture}%
}
\date{}


\author{Jean-Romain Luttringer, Quentin Bramas, Cristel Pelsser, Pascal Mérindol\\University of Strasbourg}

\maketitle
\copyrightnotice
\begin{abstract}
Interactions between the intra- and inter-domain routing protocols received little attention despite playing an important role in forwarding transit traffic. More precisely, by default, IGP distances are taken into account by BGP to select the closest exit gateway for the transit traffic (hot-potato routing). Upon an IGP update, the new best gateway may change and should be updated through the (full) re-convergence of BGP, causing superfluous BGP processing and updates in many cases. We propose OPTIC (Optimal Protection Technique for Inter-intra domain Convergence), an efficient way to assemble both protocols without losing the hot-potato property.

OPTIC pre-computes sets of gateways (BGP next-hops) shared by groups of prefixes. Such sets are guaranteed to contain the post-convergence gateway after any single IGP event for the grouped prefixes. The new optimal exits can be found through a single walk-through of each set, allowing the transit traffic to benefit from optimal BGP routes almost as soon as the IGP converges. Compared to vanilla BGP, OPTIC's structures allow it to consider a reduced number of entries: this number can be reduced by 99

\end{abstract}

\section{Introduction}
\label{secIntroduction}


\blfootnote{This work was partially supported by the French National Research Agency (ANR) project Nano-Net under contract ANR-18-CE25-0003.}

The Internet is composed of independent domains known as Autonomous Systems (ASes).
An Internal Gateway Protocol (IGP), such as OSPF or IS-IS, provides intra-domain connectivity while the Border Gateway Protocol (BGP) allows ASes to trade transit traffic.
ASes exchange routes through eBGP, while iBGP enables their dissemination among border routers.
Since several distinct routes may exist for a given BGP prefix, border routers determine the best route towards each prefix by running the BGP \textit{decision process}.
This process consists, for each prefix, in comparing routes thanks to a lexicographical order based on a set of ranked attributes (see Table~\ref{fig:bgp-ranking}).
As of 2020, the number of BGP prefixes has reached 800K~\cite{potaroo}, making this process computationally expensive~\cite{filsfils2011bgp}.

However, BGP suffers from being tightly coupled to the IGP. Between two routes equally good inter-domain wise, routers choose the one advertised by the closest gateway in terms of IGP distance (\textit{hot-potato routing}, step 6 in Table~\ref{fig:bgp-ranking}). This may affect the majority of prefixes~\cite{Agarwal04controllinghot}.
Thus, this slow decision process should be run for all prefixes by default whenever an IGP event occurs (any single internal failure or weight change, gateway included).
Since intra-domain changes are frequent~\cite{networkfailures,merindol_fine-grained_2018}, \CR{this interaction becomes a challenging
issue.}

\CR{While many fast re-routing schemes have been proposed both at the intra-~\cite{raj_survey_2007,franccois2014topology} and inter-domain (local or remote) scale~\cite{cardona_bringing_2015}, none guarantees a fast re-routing of \textit{transit traffic} towards \textit{optimal BGP routes} after \textit{any internal event}.
However such events may lead to long-lasting connectivity loss~\cite{teixeira_impact_2008, filsfils2011bgp}, performance degradation~\cite{eventimpact} and non-negligible churn~\cite{chuah_measuring_nodate-1}.}

In this paper, we present OPTIC (\textbf{Optimal Protection Technique for Inter-Intra domain Convergence}), a multi-scale routing scheme minimizing the impact of IGP changes on the BGP convergence, while enforcing both hot- and cold-potato routing.
It fulfills two design requirements, effectively making the transit traffic impervious to any IGP event.
First, transit traffic is quickly re-routed towards the new optimal BGP route \CR{in a time equivalent to the IGP convergence} (without running a full BGP decision process).
\CR{We say that OPTIC \textit{optimally protects} BGP prefixes, meaning that whatever the IGP event, the BGP prefixes are almost immediately reachable again through their best BGP route. }
Second, the background processing performed to anticipate any next future IGP event is manageable, \ie, at worst similar to BGP for the current event, but negligible when the event does not hamper the bi-connectivity of border routers.

\CR{To achieve these requirements, OPTIC efficiently computes groups of prefixes that have the same set of current and future optimal reachable gateways\footnote{Since we also consider the failure of an external gateway, the term \textit{gateway} refers to the external gateway of the neighboring AS by default (but can be limited to internal ones when the \texttt{next-hop-self} feature is used).}. After an IGP event, instead of running the BGP decision process per prefix, OPTIC's convergence relies on a simple walk-through of each group of prefixes (in practice, a simple min-search on lists of distances).
\CR{To limit the number of routing entries as well as the scale of the updates, the same set of routes is shared in memory by a group of prefixes: updating a shared set thus updates all grouped prefixes}.
Once the traffic is optimally re-routed, and only if the bi-connectivity among the set of border routers is lost, groups and their associated sets should be updated in background to \textit{anticipate} any \textit{next} future event.}
\CR{Constructing and updating such sets can be done efficiently as it does not require to consider independently each possible IGP event. With a single and simple computation, OPTIC encompasses all possible events while keeping sets at a manageable number and size.}

In Section~\ref{secRel} we describe the underlying context and
current solutions.
Section~\ref{secProblem} formally showcases the relationships between BGP
and the IGP.
Sec.~\ref{secOptic} introduces the main building blocks of our
proposal, from the data structures used to the algorithms at play.
Sec.~\ref{secResults} focuses on the performance of our protecting scheme: we analytically predict the operational cost of our enhanced data plane.
We conclude the paper with a summary of our main
achievements in Sec.~\ref{secConclusion}.

\section{Background}
\label{secRel}

\changed{
	Among previous studies enhancing BGP's convergence time,
	some suggest tuning the BGP Control-Plane through timers~\cite{pradeep2012reducing,maccari2016pop}, ghost-flushing~\cite{bremler2003improved}, modifying update messages~\cite{pei2005bgp}, using consistency assertions~\cite{cardona_bringing_2015}, or limiting path exploration~\cite{chandrashekar2005limiting}.
	These works do not prevent superfluous BGP convergence due to IGP events, neither do they allow the optimal protection of external destinations as we aim to do.
	}


	Other works reduce the impact of BGP events through data-plane anomaly detection~\cite{holterbach2019blink, holterbach2017swift, kushman2007r} to shorten reaction time. They are however focused on external events and mainly aim to isolate the data-plane from the control-plane.

	%
	Closer to our work, BGP PIC~\cite{filsfils2011bgp} aims at mitigating the effect of network failures through
	the use of a specifically designed FIB that supports backup routes (usually, a single backup route). The structure storing the optimal and the backup routes can be shared by several prefixes, reducing the update time.
	\CR{PIC relies on a hierarchical FIB, allowing the transit traffic to benefit from the IGP convergence if an internal event makes the current path to the BGP NH unusable.}
	However, as we will exhibit later on, PIC does not ensure that the transit traffic benefits from the new best BGP NH, nor the protection of the prefixes in all network configurations.

	Fast-rerouting requires improved route visibility to learn backup routes, achievable through 
	improved iBGP topologies~\cite{pelsser_improving_2008},  
	centralized architectures~\cite{gamperli2014evaluating, caesar2005design} or BGP extensions.
	In particular, BGP Add-path~\cite{simpson2014best} allows the exchange of multiple BGP routes.
	It enables fast re-routing, load-balancing, and reduces the iBGP churn.
	However, these solutions are mainly control-plane focused. They do not by themselves allow to fully benefit from the potential of the exchanged routes
	and thus do not guarantee the optimal protection of a destination.
	Nevertheless, OPTIC could be fitted on top of these control-planes (in particular Add-path with its double IGP wide option) to benefit from 
	better route visibility.

	\begin{figure}
		\centering
		\includegraphics[scale=0.93]{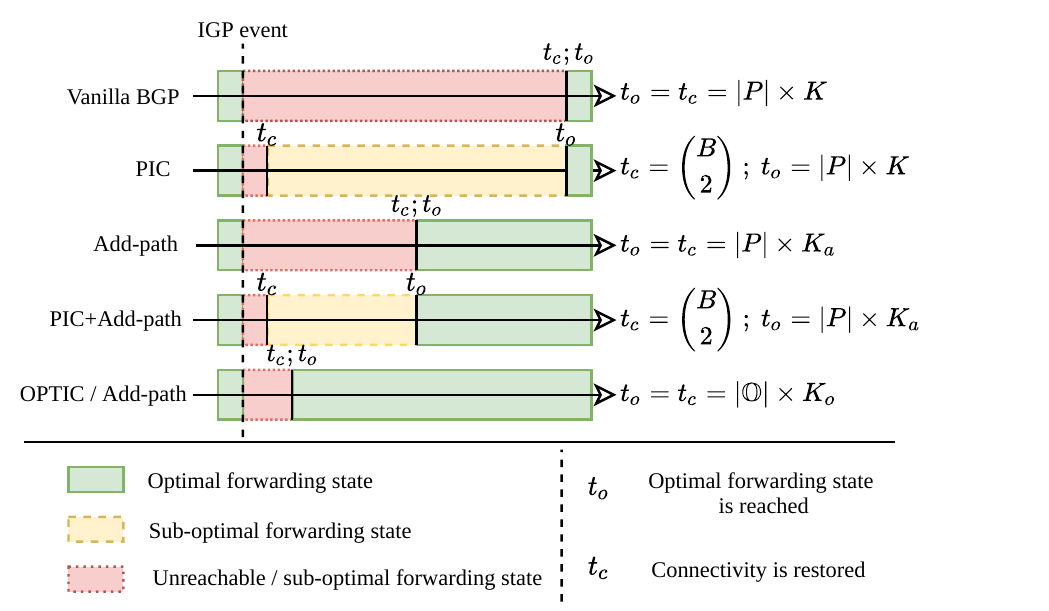}
		\caption{Connectivity and optimal forwarding state restoration timelines
		according to different technologies after internal events, depending on the number of prefixes \CR{$|P|$ and the number of BGP entries ($K_o$ when using OPTIC and $K_a$ when using Add-path)}. }
		\label{fig:timelines}
		\vspace{-2mm}

	\end{figure}

	\CR{Fig.~\ref{fig:timelines} is a pedagogical illustration that does not provide a comprehensive comparison but shows typical cases to
	position OPTIC's objectives compared to current solutions. }
	\CR{Since BGP routers only exchange their best route towards a given prefix, finding the new optimal forwarding state with vanilla BGP often require message exchange if the
	route becomes unusable. In any case, the router is required to perform a lexicographical comparison on all known routes ($K$) for each prefix ($P$).}

	\CR{PIC is designed to restore connectivity quickly by going through each of its sets
	of two gateways (whose number, for $B$ border routers, can go up to $\binom{B}{2}$)
	and falling back to the backup route of the set, or by benefiting from the IGP convergence through its hierarchical FIB.
	However, afterward, finding the new optimal gateway may still require message exchanges and a lexicographical comparison for all prefixes.
	In worst cases, the set of two gateways is not sufficient to protect the prefix (both gateways are unreachable after the event). In such cases, the connectivity cannot be restored immediately: $t_c$ can be as long as $t_o$.}

	\CR{Add-path allows to exchange subsets of routes through iBGP. With the double IGP-wide option in particular, the subsets of routes are likely to contain the new optimal gateway after any IGP event. Through adequate configuration, a BGP router can locally find the new optimal forwarding state by running the BGP decision process on the subset of routes sent through Add-path ($K_a$) for all prefixes $|P|$. This is however not fully guaranteed depending on the network connectivity. To ensure the protection of the prefixes upon any failure, all routes should be exchanged which scales poorly.}

	\CR{By combining PIC and Add-path, one can benefit from the enhanced connectivity restoration time of PIC and the advantages of Add-path. However, PIC and Add-path are not designed as a single entity and their union does not allow to reach the full potential of the available gateways. 
	While PIC can restore connectivity quickly by walking through its $\binom{B}{2}$
	sets, the time taken to restore the optimal forwarding state is ultimately the same as
	the one of Add-path alone.}


	\CR{OPTIC is designed to fully harness the potential of increased iBGP route visibility.
	By efficiently ensuring that the pre-computed sets of gateways \textit{always} possess the new optimal path whatever the network configuration, OPTIC guarantees a fast switch to the latter (thus, $t_c = t_o$) after a single walk-through of our pre-computed sets of gateways. The number ($|\mathbb{O}|$) and size ($K_o$) of these sets are both limited, as shown in our evaluation. Restoring connectivity optimally does not require to work at the prefix granularity but at the set granularity instead ($|\mathbb{O}| << |P|$ in practice).
	In some degraded cases, the sets of gateways may need to be re-computed to handle any \textit{future} IGP event (while the transit traffic already benefits from the new current optimal route). With OPTIC, this process does not rely on the slow lexicographical full BGP comparison anymore but rather on efficient updates of gateway structures and prefix groups, which remain stable when the network remains bi-connected after the change.}

\section{Keeping the potato hot}
\label{secProblem}

In this section, we show why IGP events require the re-convergence of BGP and why current solutions fail to address this challenge. Finally, we discuss how to reassemble both protocols gracefully.


\subsection{BGP/IGP: an Intimate Relationship}
\label{secDecoupling}

\begin{table}
    \centering
    
  \caption{Simplified BGP route selection}

    \begin{tabular}{@{} r L L}
        \toprule
        \emph{Step} & \emph{Criterion} & \\ \midrule
        \tikzmark[xshift=-10pt,yshift=1ex]{x}1 & Highest \texttt{local-pref} LP & (economical relationships)\\
        \rowcolor{black!5}[0pt][0pt] 2 & Shortest \texttt{as-path} & \\
        3 & Lowest \texttt{origin} & \\
        \rowcolor{black!5}[0pt][0pt] \tikzmark[xshift=-10pt,yshift=-1.2ex]{y}4 & Lowest MED & (cold-potato routing) \\
        \hline
        \hline
        \tikzmark[xshift=-10pt,yshift=1ex]{w}5 & eBGP over iBGP & \\
        \rowcolor{black!5}[0pt][0pt] 6 & Lowest IGP cost & (hot-potato routing)\\
        \tikzmark[xshift=-10pt,yshift=-1ex]{z}7 & Lowest \texttt{router-id} rid & (arbitrary tie-break)\\
        \bottomrule
        \hline
        \drawbrace[brace mirrored, thick]{x}{y}
        \drawbrace[brace mirrored, thick]{w}{z}
        \annote[left]{brace-1}{$\beta$~ }
        \annote[left]{brace-2}{$\alpha$~ }
	\end{tabular}\label{fig:bgp-ranking}
	\vspace{-2mm}
\end{table}

We start by showcasing the IGP-BGP interplay, resulting
in the need to go through all BGP prefixes after IGP events.
BGP routes are characterized by a collection of attributes of decreasing importance (Table~\ref{fig:bgp-ranking})
that can be locally modified by each router. Each attribute comes into
play whenever paths could not be
differentiated through the previous one. This route selection is
called the \textit{BGP decision process}. \CR{Note that the MED differs from the other attributes.
It can be used in different ways, but should only be used to compare routes that originate from the same AS, breaking the total order of the decision process. While our solution can be adapted to any MED flavors, we consider this most general one.}

\begin{figure}
    \centering
    \includegraphics[scale=0.45]{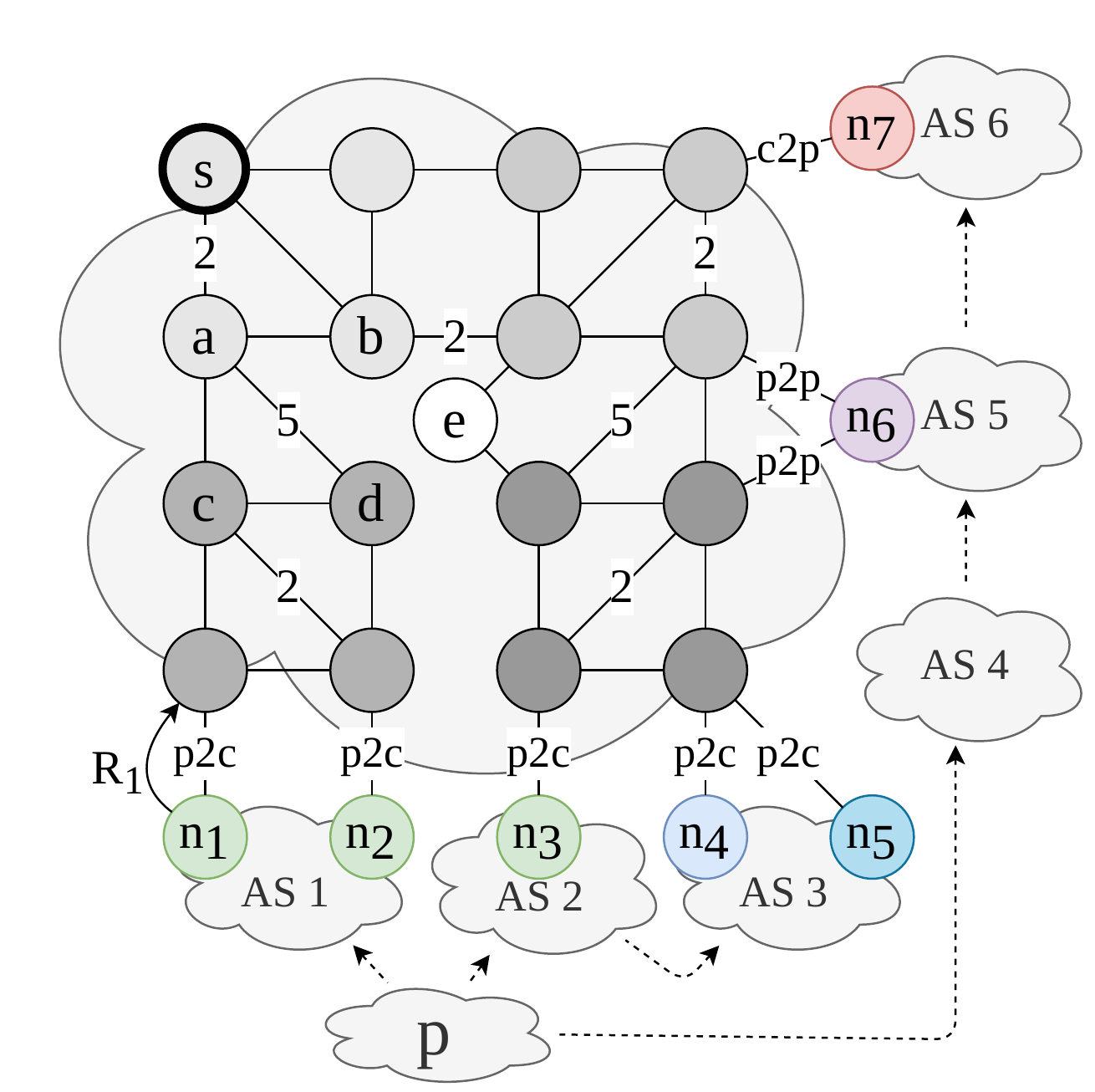}
    \caption{This example consists of an AS that learns routes towards \prefix{} via several border routers,  focusing on the point of view of s. Each link from an internal border router to the BGP NH is labeled with the type of relation between the two ASes (p2c means provider-to-customer, p2p and c2p, peer-to-peer and customer-to-provider respectively, modeled in practice by a decreasing local preference). A route R\tss{x} is advertised by the BGP NH n\tss{x}. The routes announced by n\tss{4} and n\tss{5} are discriminated through the MED attribute. Unlabeled edges weight one.}
    \label{fig:mainexample}
\end{figure}

When the inter-domain related attributes (local-pref, as-path \CR{and MED}) of two routes are equal, routers choose
the route with the closest exit point in terms of IGP distance (hot-potato routing, Line 6 in
Table~\ref{fig:bgp-ranking}).
This criterion is at the core of the
dependency between BGP and the IGP. To exhibit this interplay,
we separate the BGP attributes into two sub-lists: $\beta$ and $\alpha$,
as shown in Table~\ref{fig:bgp-ranking}. $\beta$ is composed of the
attributes purely related to inter-domain routing. They usually remain unchanged
within an AS but some operators may configure them to change in iBGP~\cite{vissicchio2014ibgp}.
Our work remains valid in both cases. Thus, for the sake of simplicity, we assume they are constant inside an AS.
The attributes $\alpha$, on the other hand, are, by construction,
router-dependent and focus on intra-domain routing.
Thus, a route \route{} towards a
prefix \prefix{} and advertised by a gateway or BGP next-hop (BGP NH) \bgpnh{}
is characterized by the vector of attributes $\beta \circ \alpha$, with $\beta
= $ [LP, as-path length, origin, med] and $\alpha = $ [ibgp/ebgp, igp
cost, router id]. Since attributes after the IGP cost are simple tie-breaks,
and since rule 5 can be seen as an extension to the IGP cost (an
eBGP route has an IGP cost of 0), we can refer to $\alpha$ simply
as the IGP distance towards the BGP NH.

It now becomes clear that IGP events may affect the ranking of BGP routes. Let us consider Fig.~\ref{fig:mainexample}. We state that \route\tss{x} \precbgp \route\tss{y} if \route\tss{x} is better than \route\tss{y} according to the BGP decision process. We
consider the routes \route\tss{1}, \route\tss{2}, \route\tss{3} and \route\tss{4} towards the prefix p announced by
\bgpnh\tss{1}, \bgpnh\tss{2}, \bgpnh\tss{3} and \bgpnh\tss{4} respectively.
\CR{The MED being irrelevant to the point, we consider that these routes have no MED for now.}
\route\tss{4} originates from a client and has an as-path length of 2, leading to the attributes $\beta(\text{\route\tss{4}}) = $ [p2c, 2, -, -]. \route\tss{1}, \route\tss{2} and \route\tss{3} are all characterized by the same $\beta(\text{\route\tss{1},\route\tss{2}, \route\tss{3}}) = $ [p2c, 1, -, -] and so are discriminated through their $\alpha$ distances (4 vs 5 vs 6).
All have a better $\beta$ than \route\tss{4}. Thus, overall, \route\tss{1} \precbgp \route\tss{2} \precbgp \route\tss{3} \precbgp \route\tss{4} from the point of view of s.

While the inequality \route\tss{1} \precbgp \route\tss{2} \precbgp \route\tss{3} holds initially,
this order is reversed after the failure of the link a$\rightarrow$c as the IGP distances, taken into account by BGP,
go from 4, 5 and 6 to 9, 8 and 6 respectively. After the failure, \route\tss{3} \precbgp \route\tss{2} \precbgp \route\tss{1}, \CR{requiring to wait for the BGP decision process to find the new best route.}
However, note that inter-domain related attributes ($\beta$) are left unaffected by an IGP event, meaning that \route\tss{4} will remain less preferred than the other three routers after \textbf{any} IGP event in any cases.

\subsection{Fast BGP re-routing upon IGP Changes}
\label{secOptProc}

We now detail why current solutions are not sufficient to guarantee fast BGP convergence.
The state-of-the-art solution would be a combination of BGP PIC~\cite{filsfils2011bgp}
(implemented on many routers) and BGP Add-Path (for path diversity).
PIC relies on the use of a Hierarchical FIB (HFIB): for each prefix, a router maintains a pointer to the BGP NH which in turn points to the IGP NH (instead of
simply memorizing the associated outgoing interface).
To protect against the failure of the gateway, PIC stores at least the two best current BGP NH, which can be learned through Add-path.
However, PIC only ensures partial sub-optimal protection and requires to run the usual BGP convergence afterward as illustrated in
Fig~\ref{fig:mainexample}.\changed{
After the failure of the link a$\rightarrow$c, PIC's HFIB restores the connectivity to \bgpnh\tss{1} by updating the IGP NH used to reach n$_1$, allowing the transit traffic to go through \bgpnh\tss{1} again.
However, this IGP event leads to a change in the $\alpha$ ranking of some BGP routes.
As seen in Sec.~\ref{secDecoupling}, after this failure, \route\tss{3} becomes the new best route.
Restoring the connectivity to \bgpnh\tss{1} without considering the changes
on $\alpha$ leads to the use of a sub-optimal route until BGP re-converges, violating so the hot-potato routing policy.
Besides, the traffic may first be re-directed after the IGP convergence, and then re-directed once again after
the BGP convergence, potentially leading to flow disruptions~\cite{disrupt}.
Letting aside the optimality issues, storing the two best BGP NH is not enough to protect the transit traffic against all failures when the network is poorly connected. Even if both \bgpnh\tss{1} and \bgpnh\tss{2} were stored, both become unreachable if $a$ fails (due to the network not being node-biconnected), leading to a loss of connectivity until BGP re-converges and finds the new best available gateway, \bgpnh\tss{3}.}
In both scenarios, retrieving the correct new optimal path requires the BGP decision process which does not scale well\footnote{
Studies proposed ways to store reduced set of routes to
enhance update times~\cite{sobrinho_distributed_2014,
thorup2001compact} but not specifically to deal with IGP changes for transit traffic.}.


\subsection{How to Reach a Symbiotic Coupling?}


We present here the necessary operational condition to untie the
BGP convergence from IGP events. The question to address is: how to efficiently pre-compute the subset composed of every BGP route that may become the new \textit{best} route upon \textit{any} IGP change?
We state that prefixes need to be \textit{optimally protected}, as per Definition~\ref{def:opt}.

\begin{mydef}{Optimal Protection}\label{def:opt}\quad \hrulefill\\
    Let \prefix{} be an external destination. We state that \prefix{} is \textbf{optimally protected} by a set \opset,
    if both pre- and post-convergence BGP NHs are stored within \opset. 
    More precisely, \opset{} should verify the two following properties for any IGP change $c$:
    \begin{itemize}
        \item (i) It contains the best BGP NH \bgpnh~towards \prefix~before $c$ occurs (pre-convergence NH for \prefix);
        \item (ii) It contains the BGP NH of the new best path
        towards \prefix{} after $c$ occurs (post-convergence NH for \prefix). It should be true for any kind of $c$: link or node event, \bgpnh{} included, such as an insertion, deletion or weight update.\quad \hrulefill
    \end{itemize}

\end{mydef}

Computing such sets naively may look costly, as predicting the optimal
gateway for each specific possible failure and weight change is time-consuming.
However, OPTIC computes and maintains these sets efficiently by \textit{rounding} them.
We will show that the size and number of such rounded sets are limited in most cases.
Finding the new optimal gateway among these sets is performed through a simple
min-search within each set (with no additional computation), updating so
each group of prefixes depending on this set, and thus every prefix.
Depending on how such sets of gateways (per group of prefixes) are designed, OPTIC can protect the traffic transiting in a BGP network against any link, node, or even SRLG \CR{(\ie, links sharing a common fate)} single failure. 

\section{OPTIC}
\label{secOptic}
%
This section explains our solution in detail.
OPTIC mitigates the impact of IGP events on the BGP convergence without hampering neither hot- or cold- potato routing. It pre-computes sets of gateways bound to
contain the current and future \textit{optimal} gateways after any single IGP failure or change, \textit{optimally protecting} every prefix\footnote{The addition of a gateway, taken into account in our set-maintaining algorithms, is  signaled by a BGP message and thus considered a BGP event.}.
For the sake of simplicity, we only consider single node and link failures (including the gateway) as well as weight changes, but OPTIC can be extended for  more general failure scenarios at the cost of complexifying the group management overhead.


\subsection{Sorting and Rounding BGP Routes}
\label{sortround}
First, we show that optimal protecting sets can be efficiently computed and maintained by sorting and \textit{rounding} BGP routes in a specific way.
We start by explaining this concept in a high-level fashion before formally detailing our solution.


\subsubsection{General idea}
\label{sec:genid}
Using our $\beta$ (inter-domain attributes) and $\alpha$ (IGP distance) attribute separation, we can compute optimal protecting sets easily.
Indeed, $\beta$ is of higher importance than $\alpha$ within the BGP decision process, and IGP events can only affect $\alpha$, leaving $\beta$ unchanged.
Thus, given the current optimal route, denoted $\route^{st}$, with $\beta(\route^{st}) = \beta^{st}$, the new optimal route after an IGP event is among the ones
with the same best $\beta^{st}$ \CR{-- we simply need to find the one with the new best $\alpha$}.
We can then easily avoid predicting which gateway will be the optimal one for
a specific event; whatever the IGP event is (except the gateway failure possibly requiring to look for more gateways), the new optimal route is among $\{\route ~|~  \beta(\route) = \beta^{st}\}$. We thus create a \textit{rounded} set that includes all routes sharing the same $\beta$.
After the IGP event, since $\beta$ attributes are unaffected, we simply need to consider that $\alpha$ may have changed and find within this rounded set the gateway with the lowest $\alpha$ (\ie, with a simple min-search).
In Fig.~\ref{fig:mainexample}, such a set would be composed of
\bgpnh\tss{1}, \bgpnh\tss{2}, \bgpnh\tss{3} as they share the same (best) $\beta$ attributes. We indeed showed in Section~\ref{secDecoupling} that any of these three gateways may become the new optimal gateway.

This is however not sufficient to deal with all failures. In particular, if the first rounded set only contains one gateway, a single failure may render all routes within the set unusable.
If this scenario occurs (because there are no two node-disjoint paths towards the external prefix -- see Section~\ref{secOinDataPlan}), more
gateways are needed to optimally protect the prefix.
Since $\beta$ attributes are unchanged by an internal event, the new best route is, a priori, among the ones with second-to-best $\beta$ attributes $\beta^{nd}$.

To form an optimal protecting set, we add rounded sets of $\beta$-tied gateways up until there is enough path diversity
to ensure that no single failure may render all of them unreachable \CR{(there are two node-disjoint paths between the border routers and prefix p)}.
By never adding less at a time than all gateways sharing the same $\beta$, we ensure that the final set contains all potential optimal gateways (as only $\alpha$ can be affected by internal events).
This final set (composed of the union of rounded sets) is thus optimally protecting, and the new optimal gateway can be found through a simple walk-through of this set after any IGP event.
If two prefixes share an equal optimal protecting set, they belong in the same group and share the same set in memory, reducing both the memory consumption and the number of entries to go through and update upon an event (as covering all shared sets covers all prefixes).
In Fig.~\ref{fig:mainexample}, \bgpnh\tss{1}, \bgpnh\tss{2}, and \bgpnh\tss{3} provided enough path-diversity to ensure the prefix was protected, \CR{and shared the same best $\beta$}. \CR{Thus, the optimal gateway after any internal event is bound to be one of these three, which then form an optimal protecting set (for all single possible failures).}

We can now present formally the data structures allowing OPTIC to compute and maintain optimal protecting sets easily.
Our solution requires to re-design both
the control- and the data-plane. The control-plane refers to all learned BGP routes. It is restructured
to ease the handling of the routes, their comparison in particular, for efficient computation of optimal protecting sets. The data-plane
only contains the information necessary for the optimal forwarding and
protection of all prefixes (\ie, the optimal protecting sets).
The resulting
structures are illustrated in Fig.~\ref{fig:optictree}, which shows how the
network depicted in Fig.~\ref{fig:mainexample} would be translated
within OPTIC's control-plane (left) and data-plane (right). \CR{To better illustrate our data structures, we assume here that n$_4$ has a better MED than n$_5$, while other routes do not possess any MED.}
The next sections describe the control-plane structure, how we construct optimal protecting sets from it, and how they are used in the data-plane.

\subsubsection{OPTIC's control-plane}
\label{secRadix}

\begin{figure}
	\hspace{-5mm}
	\centering
    \includegraphics[scale=0.55]{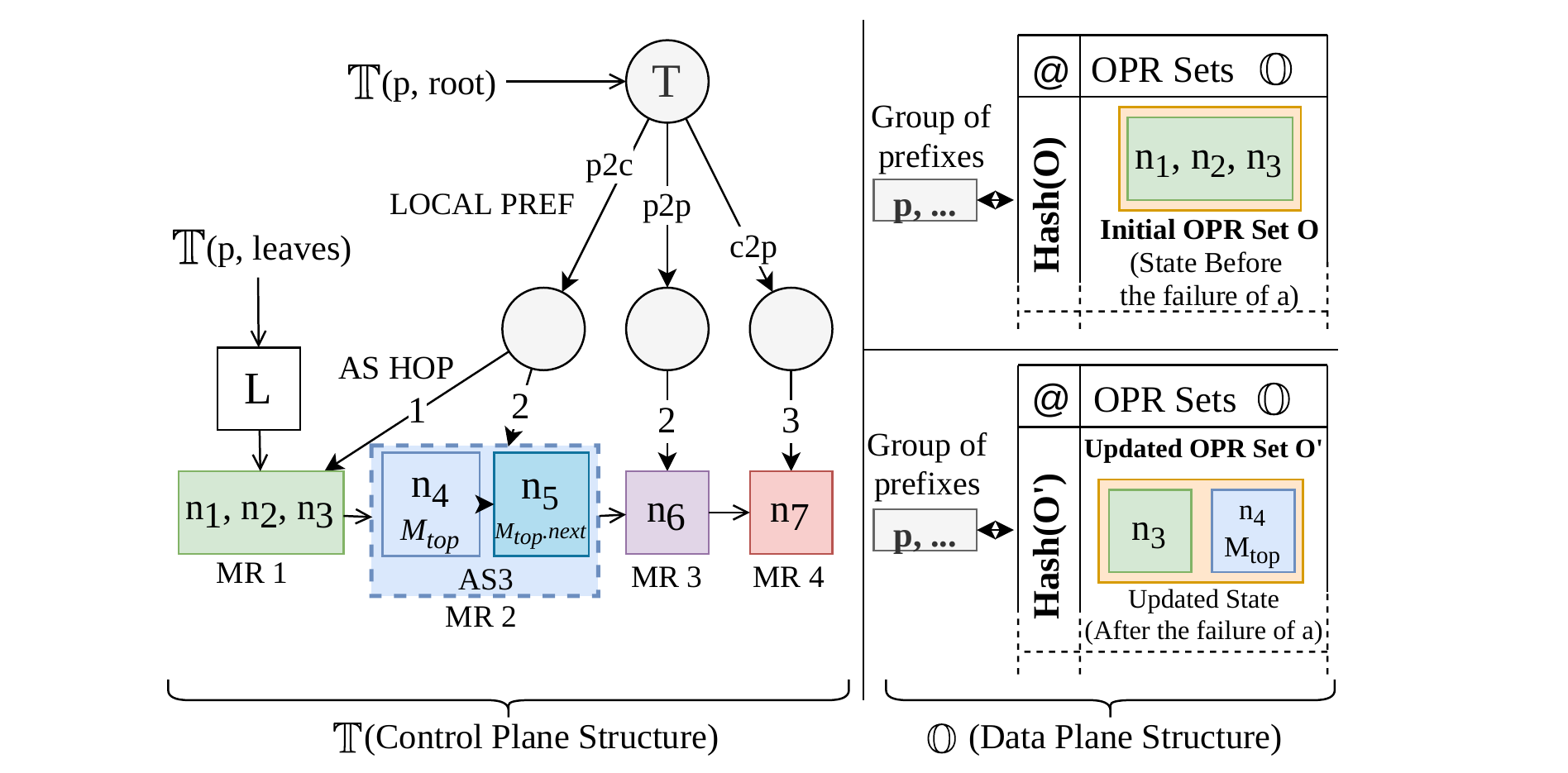}
    \caption{OPTIC's data-plane and control-plane data structures.
        In the control-plane, routes are sorted within a prefix tree $T$ whose leaves form a linked list $L$ of structured BGP NH. MED-tied routes from the same AS are chained within a linked list inside their leaf.
        Only a sufficient optimally protecting subset $O$ of routes is pushed to the data-plane.}
    \label{fig:optictree}
\end{figure}


At the control-plane level, OPTIC stores every BGP routes learned within a sorted
prefix-tree referred to as $T$, whose leaves form an ordered linked-list $L$,
which contains rounded sets of routes sorted by decreasingly preferred $\beta$ attributes.
Both $T$ and $L$ are per-prefix structures. The set of all trees, for all prefixes,
is referred to as $\mathbb{T}$.
It is important to observe that, since $\alpha$ is not considered, the tree and the list stay stable upon IGP changes and that
routes sharing the same $\beta$ attributes are stored within the same leaf.



\CR{This observation implies that when an IGP event occurs, the BGP NH of the new optimal route belongs to the first leaf of the list $L$ that contains at least one reachable gateway.}
In addition, a route from another leaf
cannot be preferred to the routes of the first leaf.
\textbf{While any route within the first leaf can become optimal after an internal change, the order of the leaves themselves can not be modified by an IGP change}.

The MED attribute can only be used to compare
routes originated from the same AS, hence we cannot use it as a global, generic
attribute. One can only consider a route with a greater MED if the
route with a better one from the same AS becomes unreachable.
Thus, routes discriminated by their MED (\textit{MED-tied} routes) for each
AS are stored within a sub-linked-list inside their leaf. This is illustrated in
Fig.~\ref{fig:optictree} with n\tss{4} and n\tss{5}. Both BGP NH share the
same three first BGP attributes and are thus stored within the same blue
leaf of $T$ (MR2). As they originated from the same AS, we store them in a sorted
linked list depending on their MED attribute.
By doing so, we consider only the first route in the \textit{MED-tied} list that is reachable (referred to as $M_{top}$), respecting the MED's semantics. \CR{Peculiar situations, such as routes not having a MED while others do, can be resolved by applying the standard ISP practices (\eg, ignoring such routes or using a default MED value)}.
The leaves of the tree $T$ thus form a sequence of rounded sets of gateways
\CR{stable upon IGP changes}. We call each set a \textit{MED-aware Rounded set}.

\begin{mydef}\label{def:RS} MED-Aware Rounded sets (MR)\quad \hrulefill\\
    For a given prefix, a leaf of its prefix tree $T$ is called a MED-Aware Rounded set. In particular, it contains all the routes having the same $\beta$ attributes (MED-Excluded).\quad \hrulefill
\end{mydef}

\subsubsection{Getting optimal-protecting sets from the control-plane}
\label{secModel}
We now explain how this construct eases the computation of optimal protecting sets.

For each prefix p, the first MR set contains, by construction, the optimal
pre-convergence route. As stated in Section~\ref{secRadix}, any BGP NH within the same MR set may offer the new optimal post-convergence route after an internal event.
However, this first MR set is not always sufficient to protect p.
In this case, the new optimal BGP
NH is bound to be within the first MR set in $L$ which contains a gateway that is still reachable.
Consequently, OPTIC constructs an optimally protecting set by considering the union of the best MR sets, in order,
until the destination prefix p is protected from any failure, \ie, there exist two node-disjoint paths. The union of such MR
sets is referred to as an Optimal-Protecting Rounded (OPR) set for p.
The formal definition is given in Theorem.~\ref{th:OPR}. Due to lack of space and its intuitive design, its proof is not presented in this paper (but available in \cite{masterthesis}).

\begin{myth}\label{th:OPR} Optimal-Protecting Rounded sets (\textbf{OPR})\quad \hrulefill\\ 
    Let $p$ be a prefix, and $M_1,M_2,\ldots$ be the sequence of MR sets in the list $L$.
    Let $\opset = \bigcup_{i=1}^{x}M_{i}$ with $x$ minimal such that there exist two node-disjoint
    paths towards $p$ (passing through $\opset$).

    \textbf{Then, $\opset$, called the Optimal-Protecting Rounded set of $p$, is optimally protecting $p$}.  \quad \hrulefill
\end{myth}

Adding MR sets until the
prefix p is protected means that there now exists
enough path diversity to protect p from any single event.
The number of routes necessary to protect a prefix depends on the resilience of the network. In bi-connected networks, two gateways are enough.



OPR sets computation does not require any (prior) knowledge of the IGP graph to cover
all possible IGP events. Verifying the existence of two node-disjoint paths between the border router and p via $\opset$ is enough and the lightest possible processing to test the protection property. Unless the protection property
is affected by the event, OPR sets stay stable.

\subsubsection{Using OPR sets in the data-plane}
\label{secOinDataPlan}
Once OPR sets are extracted from the control-plane, we push them to the data-plane.
The bottom part of
Fig.~\ref{fig:optictree} shows OPTIC's data-plane.
For a given prefix, only the OPR set O (and not the whole list $L$) that optimally protects p is pushed to the data-plane. The data-plane contains the meta-set \opsets{} of all OPR sets for all groups of prefixes,
indexed by their hash, as shown in Fig.~\ref{fig:optictree}.
Prefixes, when sharing the same OPR set, point towards the same set O. 
The hash index is content-based (see next sections for more details) and eases the management of \opsets{}. Allowing prefixes to share the same O reduces the amount of data that has to be
stored within the data-plane, as well as the scale of the updates. Note that,
since O is constructed from a subset of $L$, prefixes can share the same OPR set O while having different control-plane structures $L$.



\begin{algorithm}\caption{update\_OPR}\label{alg:upOPR}
    \footnotesize
    \hrulefill
    \SetAlgoLined
    \SetAlgoVlined

    \KwData{ \leaveslist, \opsets, oldH, \prefix,  \igpgraph, \igpdists
    }

    \KwResult{ updates OPR sets and \pfxsetbgp}
    \vspace{-2mm}
    \hrulefill

    \SetKwProg{Fn}{Function}{ $\rightarrow$}{end}
    \Fn{update\_OPR}{
        \nl \opset~ = \kwEOS(\leaveslist, \igpgraph)\; \label{alg:uOR:extract}
        \nl \For{$M_{top} \in \opset$}{
                            \nl\While{M$_{top}$ $\neq \emptyset$\label{alg:uOR:reachable?}}{
                                \nl M$_{top}$.\igpcost{} = \igpdists[M$_{top}$]\;
                                \nl M$_{top}$ = M$_{top}$.next\; \label{alg:uOR:med}
                            }
        }

        \nl        \opset$_{top}$ = \textbf{min}$_\alpha$(\opset)\; \label{alg:uOR:topg}

        \nl newH = \hash(\opset)\;
        \nl     \If{\opset~$\notin$ \opsets}{
            \nl         \opsets[newH]~= \opset\; \label{alg:uOR:insert}
        }
                \nl         \pfxsetbgp(\prefix) = newH\;
            %

        \nl        \If{$\not\exists$~ \prefix~ $|$ \pfxsetbgp(\prefix) = oldH}{
            \nl            \remove \opsets[oldH] \from \opsets;\label{alg:uOR:remove}
        }
    }

	\hrulefill\vspace{2mm}
\end{algorithm}

\changed{
Algorithm~\ref{alg:upOPR} shows how the OPR sets are updated  in the data-plane when necessary. The optimal protection property may require to add gateways from the data-plane structure \opset{} (while removals are performed for efficiency).
We start by extracting the OPR set \opset{} from the control-plane structure \leaveslist{} (Line~\ref{alg:uOR:extract}).
The required MR sets are computed by first \emph{(i)} creating a graph $G'$ from $G$ where we add a virtual node representing a remote prefix, then \emph{(ii)} connecting in $G'$ the gateways from MR sets, MR per MR to this virtual node, until there exist two node-disjoint paths towards the virtual node.
\textbf{ExtractOPRSet} thus returns an OPR set as defined by Theorem~\ref{th:OPR}. We then add the IGP distances towards each gateway (contained within $D$)
 to the structure (Line~\ref{alg:uOR:reachable?}). This is done for each gateway, including MED-tied ones (Line~\ref{alg:uOR:med}).
}
Finally, OPTIC retrieves the current optimal gateway within O, \opset$_{top}$, \ie, the one with the lowest IGP distance (Line~\ref{alg:uOR:topg}).

Once the OPR set O is updated, we compute its hash to check its existence within \opsets{} and
insert it if required (Line~\ref{alg:uOR:insert}).
Finally, if no prefixes still use \CR{the previous} O, it is removed from the data-plane (Line~\ref{alg:uOR:remove}).
This algorithm maintains the data-plane in an optimal-protecting state.
Its limited complexity is often bypassed (after the bootstrap), as we expect
OPR sets to stay stable in bi-connected networks. The complexity of \textbf{extractOPR} scales linearly in the IGP dimensions.
Unused OPR sets could be kept transiently to mitigate the effects of intermittent failures.



\subsection{Dealing with BGP and IGP events}
\label{secAlgs}

We describe here how OPR sets are
 updated upon a BGP or an IGP event to achieve
optimal protection of all destinations.



\begin{algorithm}\caption{BGP\_Update}\label{alg:bgp}
	\footnotesize
    \hrulefill
    \SetAlgoLined
    \SetAlgoVlined

    \KwData{ \rdxtreeset, \opsets, \pfxsetbgp, \route~=~(\prefix, \bgpnh, $\beta$), \igpgraph, \igpdists}

    \KwResult{ Update of \rdxtreeset, \opsets, \pfxsetbgp\\}
    \vspace{-2mm}
    \hrulefill

    %

    \SetKwProg{Fn}{Function}{ $\rightarrow$}{end}
    \Fn{BGP\_Update}{
        \nl \rdxtree~ = \rdxtreeset(\prefix, tree)\; \label{alg:bgp:getT}
                \nl H = \pfxsetbgp(\prefix); \tcp{H = \hash(\opset)} \label{alg:bgp:hash}
        \nl \leaveslist~ = \rdxtreeset(\prefix, leaves)\; \label{alg:bgp:getL}
        \nl \uIf{Event = Add}{
            \nl rMR = \add \route~\kwin \rdxtree \; \label{alg:bgp:add}
        }
        \nl \Else{
            \nl rMR = \remove \route~ \from \rdxtree\; \label{alg:bgp:remove}
        }

         \nl \If{rMR $\in$ \kwusefulMR(\leaveslist)}{\label{alg:bgp:useful}
            \nl     \upOP(\leaveslist, \opsets, H, \prefix, \igpgraph, \igpdists)\;\label{alg:bgp:recomp1}
                            }
    }
    \hrulefill\vspace{2mm}
\end{algorithm}

\subsubsection{BGP updates}

Algorithm~\ref{alg:bgp} showcases how to maintain OPR sets upon a BGP update, being either an \textit{Add} (\ie, a new route is learned) or a \textit{Down} event (\ie, a
withdraw that cancels a route)\footnote{A modified route can
be handled through a \textit{Down} followed by an \textit{Add}.}.
As a BGP update concerns a
given prefix, only one OPR set O (the one that optimally protects p) is modified when necessary. Intuitively,
checking whether the route R belongs (or should belong) to the leaves of $T$ extracted to create the current O (\ie, if R belongs to the current O) is enough to know if the update is necessary.

First, Alg.~\ref{alg:bgp} retrieves the route-tree $T$
of the updated prefix p (Line~\ref{alg:bgp:getT}).
Depending on the nature of the update,
we update the control-plane structure $T$ (and implicitly $L$) by
either adding (Line~\ref{alg:bgp:add}) or removing
(Line~\ref{alg:bgp:remove}) the updated route.
When performing these operations, we store the rank of the MR set containing the route R, \textit{rMR}.

%


Using rMR, one can check whether R belongs (or should belong) to O (Line~\ref{alg:bgp:useful}), \eg, by memorizing the number of MR sets used to form O.
If R is not good enough to belong to the current OPR set, there is no need
to consider it and the algorithm ends.
Otherwise, if R is a newly added (resp. withdrawn) route, it must be added
(resp. removed) from the data-plane structure \opset{} which can be found in $\mathbb{O}$ through its hash.
In both cases, \opset{} has to be updated (Line~\ref{alg:bgp:recomp1}).
One can see that OPTIC's behavior when dealing with a BGP update is pretty
straightforward and that  BGP events are likely to have no bearing on the data-plane.


\begin{algorithm}\caption{IGP\_Change}\label{alg:igpup}
    \hrulefill
    \SetAlgoLined
	\SetAlgoVlined
	\footnotesize

    \KwData{ \rdxtreeset, \opsets, \pfxsetbgp, \igpgraph$=(E,V)$, \link, \weight }

    \KwResult{ Update of \rdxtreeset, \opsets, \pfxsetbgp, \igpgraph\\}
    \vspace{-2mm}
    \hrulefill

    \SetKwProg{Fn}{Function}{ $\rightarrow$}{end}
    \Fn{Change}{
        \nl \igpdists~ = \spf(\igpgraph, \link, \weight)\;
        \nl \ForAll{\opset~ $\in$ \opsets}{ \label{alg:igp:allO}
            \nl\ForAll{M$_{top}$ $\in$ \opset}{\label{alg:igp:through}
                                \nl\While{M$_{top}$ $\neq \emptyset$ $\wedge$ \igpdists[M$_{top}$] $= \infty$\label{alg:igp:reachable??}}{
                                    \nl M$_{top}$ = M$_{top}$.next\; \label{alg:igp:med}
                                }
                                \nl\uIf{M$_{top}$ $= \emptyset$}{
                                    \nl \remove M$_{top}$~ \from \opset\;  \label{alg:igp:remove}
                                }
                                \nl\Else{
                    \nl                M$_{top}$.$\alpha$ = \igpdists[M$_{top}$]\;\label{alg:igp:weight}
                }
            }
            \nl \opset$_{top}$ = \textbf{min}$_\alpha$(\opset)\; \label{alg:igp:opt}


            \nl\uIf{\link~ $\in E \wedge~ w \neq \infty$}{ \label{alg:igp:cond2}
                \nl \kwcontinue\;
            }
            \nl\ElseIf{\ispro(\opset, \igpgraph) $\wedge~ \kwismin(\opset)$}{{\label{alg:igp:test2disjoint_routes}
                  \nl \kwcontinue\;
                }
            }

            \nl \add \hash(\opset) \kwin OPR\_to\_update\;
        }
        \nl\ForAll{$H \in OPR\_to\_update$}{\label{alg:igp:broke}
        \nl         \ForAll{$\prefix~ |~ \pfxsetbgp(\prefix) =$ H}{
            \nl\leaveslist~ = \rdxtreeset(\prefix, list)\;
            \nl\upOP(\leaveslist, \opsets, H, \prefix, G, \igpdists)\;
        }
        }
    }

    \hrulefill\vspace{2mm}
\end{algorithm}

\subsubsection{IGP changes}

Algorithm~\ref{alg:igpup} showcases the behavior of OPTIC upon an IGP
change, which can be a modification on the existence (insertion or
deletion -- modeled by an infinite weight) or on the weight \weight{} of a link \link{} (a node wide change can be modeled through its multiple outgoing links).

Upon a change, the new IGP distances \igpdists{} are
recovered. OPTIC then considers each \opset, covering so every BGP prefixes
(Line~\ref{alg:igp:allO}).
For each relevant gateway (with the best MED for each AS, $M_{top}$) within \opset, we first check
whether it is still reachable (Line~\ref{alg:igp:reachable??}). Unreachable
gateways are replaced by the next MED-tied route when possible (Line~\ref{alg:igp:med}) or removed (Line~\ref{alg:igp:remove}) otherwise. Reachable gateways are
first updated with their new best IGP distances (Line~\ref{alg:igp:weight}).
\changed{The whole group of prefixes using the set benefits from its new optimal path  as soon as possible (Line~\ref{alg:igp:opt}).
Afterward, if necessary, we update OPTIC's structures in the background to anticipate any future internal event.}

If the updated link \link{} is a
valuation change (Line~\ref{alg:igp:cond2}), there is no loss of reachability. Thus, \opset{} still contains two disjoint paths towards p and remains stable.
For other kinds of events, \opset{} may need to be updated, as
connectivity may have evolved due to the insertion or deletion of a link.
If a link was added, the network connectivity may have increased and useless MR sets can be removed
if \opset{} is not already minimal (\eg, containing two gateways).
If a link was removed, \opset{} may have lost its protecting property
and may have to be updated. These two scenarios, leading to the update of
\opset{}, are visible in the condition Line~\ref{alg:igp:test2disjoint_routes}.
This update is used to prepare for a future event.
We perform it in background afterwards (Line~\ref{alg:igp:broke}) and continue to walk through \opsets{} to restore the optimal forwarding state of all groups of prefixes quickly.


The update aforementioned (Line~\ref{alg:igp:broke}) is performed at the prefix granularity (\ie, for each prefixes that used \opset{} that will be updated).
Indeed, while these prefixes share the same O before the change,
they do not necessarily share the same $L$. Since O may be updated by fetching
information from $L$, they may point to distinct OPR sets after the update.
Recall that this is a background processing phase where OPTIC may fallback to the prefix granularity to anticipate the next change only if node-bi-connectivity is not granted anymore.
The fast switch to the new optimal post-convergence gateway was already performed at Line~\ref{alg:igp:opt}.
This switch is not done at the prefix granularity, it is performed only for each \opset{} instead.

In short, most BGP and IGP updates both result in simple operations.
A BGP update just triggers a prefix tree manipulation: a single OPR set is re-computed only if the updated route is, or should be, part of the set.
An IGP weight-change 
results in the walk-through of all OPR sets ($\opsets$) \CR{and a min-search} to converge to the new optimal forwarding state \CR{followed by a background processing if necessary}.
We argue that the cardinal of $\opsets$ will be orders of magnitudes lower than the number of BGP prefixes in most networks.
The failure or addition of a link or node results in the same walk-through, but could also require the background update of some OPR sets to prepare for a future event.
More precisely, only when the network gains or loses its bi-connected property could some OPR sets be affected. New OPR sets then need to be re-computed for the prefixes of the groups that depended on the affected OPR sets. 
Instead of the number of prefixes, OPTIC convergence scales with the number and the size of the OPR sets.
Consequently, to assess the viability of OPTIC, we aim at limiting
their size (and so number). While Sec.~\ref{secResults} analyzes that aspect in
detail, the next subsection explores conditions on the graph properties allowing to use smaller optimally protecting sets.

\subsection{Optimizations}
\label{ssec:opti}

In this section, we introduce some
optimizations that allow to reduce the size of the OPR sets used by OPTIC.

Let us start with a fairly reasonable assumption: well-designed networks should
offer bi-connectivity between border routers.
Based on this realistic hypothesis, we can consider two kinds of
reductions: $(i)$ removing MED-tied entries from an OPR set and $(ii)$ discarding
all gateways in the second MR set except the best one (when the first MR set includes only one gateway).
As the first optimization will not be mentioned further on, we will not dwell on it.
Intuitively, since the MED attribute is of higher importance than the IGP cost,
it may allow us to remove routes with lower MED from the set, as an IGP cost change will not make these routes optimal.


This second optimization will be evaluated in our theoretical analysis and allows
to keep at most one gateway from the second MR set when the first one contains a single gateway.
If the current optimal gateway is not part of the path towards the first gateway of the second MR set, adding this
second gateway is enough to form an OPR set. When the first MR set is made of only one gateway and the network bi-connected, OPTIC only needs to consider other gateways for the specific case of the optimal gateway failure (as other changes cannot make it less preferred than gateways from the following MR sets).
If the second-best gateway does not use the first to reach p, its IGP
distance will not be impacted by the current gateways' failure, and it will become, after the failure,
the best gateway overall. This allows OPTIC to create many OPR sets containing only two routes.

\section{Data Plane Scalability Analysis}
\label{secResults}
To react to an IGP event, OPTIC only operates a min-search in all
OPR sets. 
 OPTIC's performances thus mainly depend on the number of OPR sets ($|\mathbb{O}|$) and their sizes.

We present here a theoretical model capturing a wide variety of scenarios.
This analytical approach is more suitable than experiments,
as it is more general and provides a pessimistic order of magnitude of OPTIC's potential. This approach yields the same results as a simulation, but allows to easily explore numerous scenarios.
It highlights what an ISP can expect by running OPTIC given only a few structural parameters on their networks.
We investigate several ASes profiles (constructed from \cite{luckie_as_2013}'s data), varying the number of gateways, peers, clients, and providers, as well as the number of prefixes learned through each of the latter. We show that $|\mathbb{O}|$ remains manageable and/or close to the lower bound, being 99\% smaller than the number of prefixes for stub networks.

\subsection{Preliminary Model: counting \#OPR sets}
\label{secComplexity}

\newif\ifmodellong

\modellongfalse

\newif\ifmodelcourt
\modelcourttrue
\ifmodellong
\modelcourtfalse
\fi

We consider an AS (or a portion of it), with $B$ bi-connected gateways advertising $P$ prefixes in total.
Each prefix \prefix{} is advertised by a subset of $b \leq B$ of those gateways, chosen uniformly at random.
The $\beta$ of each prefix is represented by a value between $1$ and $ps$ (policy spreading) also chosen uniformly at random.
For a given \prefix, this implies that any subset of gateways of a given size $n \leq b$ all have the same probability to be the OPR set for \prefix.
Our model analyzes the number $|\opsets| = |{\opsets}_{B,P,ps}|$ of unique OPR sets depending on the number $B$ of gateways, the number $P$ of prefixes, and on the policy spreading $ps$. In practice, we decide to set $b$ to a constant value (\eg, $b=5$) greater than the median in \cite{luckie_as_2013}.

The quantity $|{\opsets}_{B, P,ps}|$ can be seen as the sum of distinct OPR sets of different sizes.
\ifmodelcourt
From our assumptions, each OPR set of size $n$ (${2 \leq n \leq b}$) is in ${\opsets}_{B, P,ps}$ with the same probability, that we denote $\P_{B, P, ps, n}$. Since there are $\binom{B}{n}$ such possible sets of size $n$, we have:
\begin{equation}\label{eq:simple model main formula}
|{\opsets}_{B,P,ps}| = \sum_{n=2}^{b} \binom{B}{n} \P_{B, P, ps, n}.
\end{equation}
For a given prefix, let $p_n$ be the probability that the size of its OPR set is $n$. We obtain
$
{\P_{B, P, ps, n} = 1-\left(1-\binom{B}{n}^{-1}\right)^{p_nP}}
$.
Since we assume that each prefix is learned by $b$ gateways, and the weight associated with each gateway is chosen uniformly at random between $1$ and $ps$, we use ``Balls into bins'' analysis to compute $p_n$:
\[
p_n = \sum\limits_{i=1}^{ps}
\frac{1}{ps^{n}}\left(1-\frac{i}{ps}\right)^{b-n}
\left(
(i-1)b\binom{b-1}{n-1} + \binom{b}{n}
\right)
\]
\fi

\ifmodellong
From our assumptions, OPR sets of size $n$ (${2 \leq n \leq b}$) are in ${\opsets}_{B, P,ps}$ with the same probability, that we denote $\P_{B, P, ps, n}$.
In other words, $\P_{B, P, ps, n}$ is the probability for a particular subset of gateways of size $n$ to be the OPR set of at least one prefix (among all the $P$ prefixes). Since there are $\binom{B}{n}$ such possible sets of size $n$, we have:
\begin{equation}\label{eq:simple model main formula}
|{\opsets}_{B,P,ps}| = \sum_{n=2}^{b} \binom{B}{n} \P_{B, P, ps, n}.
\end{equation}

For a given prefix, let $p_n$ be the probability that the size of its OPR set is $n$. Hence, the number of prefixes associated with OPR sets of size $n$ is $p_nP$ in average. Thus, the probability that a given OPR set of size $n$ is not associated with any prefix is $\left(1-\binom{B}{n}^{-1}\right)^{p_nP}$. So, the probability that a given OPR set is associated with at least one prefix is
$
{\P_{B, ps, P, n} = 1-\left(1-\binom{B}{n}^{-1}\right)^{p_nP}}
$.

We now compute the probability $p_n$ for a given prefix to have an OPR set of size $n$.
Since we assume that each prefix is learned by $b$ gateways, and the weight associated with each gateway is chosen uniformly at random between $1$ and $ps$, we can use ``Balls into bins'' techniques to count how many gateways have minimal weight and to obtain \emph{(i)} the probability that only one or two gateways have minimal weight, in which case the OPR set is of size 2, thanks to our optimization and \emph{(ii)} the probability that $n \geq 2$ gateways have minimal weight, in which case the OPR set is of size $n$:

\[\scriptstyle
p_n=\left\{
\begin{array}{ll}
\sum\limits_{i=1}^{ps}\frac{b}{ps}\left(1-\frac{i}{ps}\right)^{b-1} + \binom{b}{2}\frac{1}{ps^{2}}\left(1-\frac{i}{ps}\right)^{b-2}&\text{if }n{=}2\\
\sum\limits_{i=1}^{ps}\binom{b}{n}\frac{1}{ps^{n}}\left(1-\frac{i}{ps}\right)^{b-n} &\text{if }n{\geq}3
\end{array}
\right.
\]
\fi

A similar reasoning can be done for our optimization looking for sets having 2 gateways.
We now take into account the specificity of the local-pref attribute.

\subsection{Towards a Realistic Evaluation}
\label{secAnalysis}


\begin{table*}
    \centering

    \caption{Number of distinct OPR sets ($|\mathbb{O}|$) for several scenarios.}
 
    \begin{tabular}{|c|c|c|c|c|c|}
        	\hline
        	AS Type & \# gw per class & \# pfx per class & \# distinct OPR sets & O median size & Lower bound \\\hline
            Stub & $(10; 20; \num{0})$  & $(700K; 100K; 0K)$ & $\num{3475}$ & $\num{4}$ & \num{235}\\\hline
Tier 4 & $(10; 25; \num{25})$  & $(500K; 200K; 100K)$ & $\num{10589}$ & $\num{3}$ & \num{645}\\\hline
Tier 3 & $(10; 50; \num{100})$  & $(500K; 200K; 100K)$ & $\num{33610}$ & $\num{3}$ & \num{6219}\\\hline
Large Tier 3 & $(10; 100; \num{500})$  & $(500K; 200K; 100K)$ & $\num{101997}$ & $\num{2}$ & \num{73781}\\\hline
Tier 2 & $(5; 500; \num{2000})$  & $(500K; 200K; 100K)$ & $\num{215429}$ & $\num{2}$ & \num{197194}\\\hline
Tier 1 & $(0; 50; \num{5000})$  & $(0K; 600K; 200K)$ & $\num{228898}$ & $\num{2}$ & \num{199633}\\\hline
\end{tabular}

  \label{tab:number of OPR sets per category}
\end{table*}

\subsubsection{Break Down Into Classes}
In practice, neighboring ASes are partitioned in several classes (\emph{eg.}, clients, peers, and providers), usually represented by the local-pref attribute.
At the end of the decision process, we know that a prefix \prefix~is associated with a single class. 
Indeed, the local-pref depends only on the set of advertising neighbors for \prefix: it belongs to the class of the neighbor having the highest local-pref.

This allows us to split the analysis by class.
With this assumption, OPR sets are included inside a unique class of gateways, but as a counterpart, the policy spreading in each class is reduced (because gateways have the same local-pref inside a class).
We use our former model to compute the number of distinct OPR sets in each class with $ps=b=5$.
This calibration is pessimistic enough as it only takes into account a limited AS length dispersion and always 5 learning gateways in the best class.


$B_1$, $B_2$ and $B_3$, denote respectively the number of gateways with local-pref 1, 2 and 3.
Similarly, $P_1$, $P_2$ and $P_3$, denote respectively the number of prefixes originating from a gateway with local-pref 1, 2 and 3.
We have $P_1 + P_2 + P_3 = P =\num{800000}$ and $B_1 + B_2 + B_3 = B$.
We can now compute $|\opsets|$ by assuming each class follows our basic model:
\[|\opsets| = |{\opsets}_{B_1,P_1,5}| + |{\opsets}_{B_2, P_2,5}| + |{\opsets}_{B_3, P_3, 5}|
\]

This sum gives the theoretical performance of OPTIC as it is the number of OPR sets each router has to manage.

\subsubsection{Definition of the Lower Bound}
We define here the best theoretical performance an optimally protecting scheme could reach, to compare it with OPTIC. Such a scheme would have to store sets of at minima two gateways (less can not ensure protection).
This lower bound also provides an estimation of the performances of techniques just aiming at providing (non-optimal) protection like~\cite{filsfils2011bgp}.
In other words, with $P_i$ prefixes and $B_i$ gateways in a given class, the average minimum number of optimally protecting sets is the average number of distinct sets obtained when choosing $P_i$ random subsets of two gateways (such sets are chosen uniformly at random). 


\subsubsection{Evaluation on Fixed Break Down}

We now compute $|\opsets|$ for several AS categories; a \emph{Stub} has few peers and even fewer providers from where most the prefixes originate; a \emph{Transit} (Tier 2, 3 and 4) has a limited number of providers but from where the majority of the prefixes originate, more peers and possibly numerous customers; a \emph{Tier 1} has few peers and a large number of customers. For \emph{Transit} and \emph{Tier 1}, we present different class and prefix break down.
Note that our model is pessimistic, as, for \emph{Tier 1} in particular, ASes may have more classes with $ps > 5$ (\eg, gateways can be geographically grouped).
The number of gateways, and their partition into classes, are rounded upper bounds of realistic values obtained from~\cite{luckie_as_2013}.
Moreover, we did not assume any specific popularity of certain gateways.
Using our complementary material~\cite{zenodo}, $|\opsets|$ can be computed for any parameters.

Table~\ref{tab:number of OPR sets per category} shows that the number of OPR sets is more than reasonable for Stubs and small Transit. For large transit, the distribution of the prefixes into classes has a great impact on $|\opsets|$.
As expected, for Tier 1, the number of OPR sets is high, but OPTIC is close to the lower bound (there is not much room for possible improvements).
The number of routes contained within each
OPR set is limited, meaning that the min-search applied upon
an IGP event has a limited computational cost.
Finally, it is worth recalling that our analysis is pessimistic because uniform. Regional preferences or gateway popularity can strongly reduce the size of \opsets{} in practice.

\newcommand{\ratio}{\delta}
\subsubsection{Evaluation on Variable Break Down}
In addition to previous specific cases, we here show how $|\opsets|$ evolves depending on the sizes of the classes and their relation.
For that, we introduce the variable $\ratio$ that represents the ratio between the sizes of two successive classes.
More precisely, when a break down considers a ratio of $\ratio$, then it means $(B_1, B_2, B_3) = (B_1,B_1\times \ratio,B_1\times \ratio^2)$.
Similarly, we also assume that the number of prefixes learned by each class verifies $(P_1, P_2, P_3) = (P_1,P_1/\ratio,P_1/ \ratio^2)$.

\begin{figure}
  \centering

    \includegraphics[width=6cm]{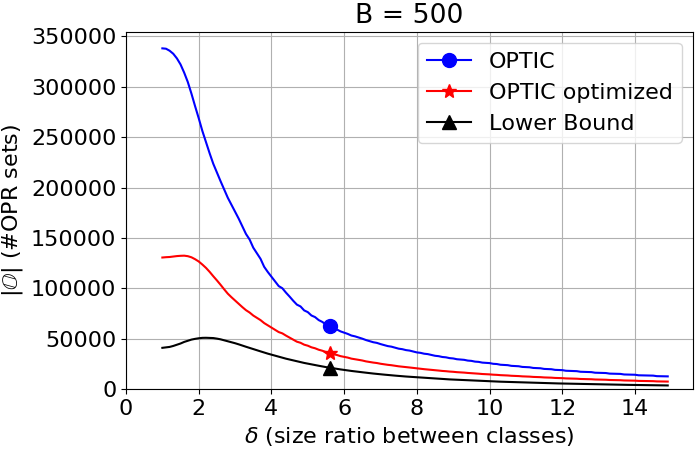}
    \caption{$|\opsets|$ depending on the ratio $\ratio$ between classes with $B=500$. 
    }\label{fig:memory usage per ratio}
\end{figure}

We present in Figure~\ref{fig:memory usage per ratio} the number of distinct OPR sets depending on $\ratio$ for $B=500$; $\ratio$ varies from 1 (all the classes have the same size) to 15 (each class has 15 times more gateways that the previous class, but learns 15 times fewer prefixes).
When the number of gateways is low, the management cost is obviously limited (even when all the classes have the same size).
With $B = 500$, OPTIC's performance is limited for small $\ratio$ but gets better quickly.
When
$\ratio \geq 8$, our optimization performs as well as the lower bound. 

\begin{figure}
  \centering

    \includegraphics[width=6cm]{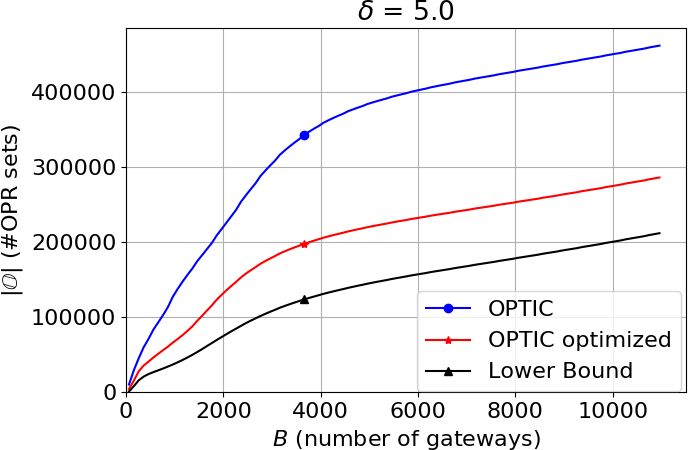}
    \caption{$|\opsets|$ depending on the number of gateways, with $\ratio = 5$.
    }\label{fig:memory usage per number of gateways}
\end{figure}

We now investigate the case where $\delta$ equals 5 and look at how $|\opsets|$ evolves depending on $B$.
We see in Fig.~\ref{fig:memory usage per number of gateways} that our optimized OPR reduction outperforms the non-optimized one.
For less than 1500 gateways, the number of OPR sets is smaller than \num{100000} with our optimized version. Then, $|\opsets|$ increases quickly to reach \num{200000} when there are around 4000 gateways, with a lower bound at \num{125000} sets. After, the growth is linear, so the proportional overhead of our solution compared to the lower bound tends to one.

The management cost of OPTIC is reasonable, especially for networks having a limited number of border gateways where OPTIC exhibits very good performances.
For large networks having numerous gateways, the size of our data-plane structures remains limited regarding the number of prefixes, and there does not exist many room for theoretical improvements.

\section{Conclusion}
\label{secConclusion}
Because the IGP and BGP are entangled to enforce hot-potato routing at the AS scale, an IGP change triggers the full and slow BGP convergence.
With OPTIC, we aim at re-designing this IGP/BGP coupling. 
We proposed efficient MED-aware algorithms and data-structures to anticipate and quickly react to any single IGP event (weight change, link or node failure, including the outage of BGP border routers).
At the data-plane level, OPTIC ensures a fast and optimal re-convergence of the transit traffic.
In the control-plane, OPTIC updates its constructs in background to anticipate a future event when necessary (only after changes modifying the 2-node-connectivity network property).

Since nearly all calculations are performed per group of prefixes, OPTIC scales orders of magnitudes lower than the number of BGP prefixes.
Our analytical evaluation shows that the number of entries to manage in the FIB is at worst 50\% of the full Internet table for large Tier-1.
It scales down to 25\% for large Tier-2s and less than 1\% for Stub AS, which represents 84\% of all ASes in the current Internet.


\bibliographystyle{IEEEtran}
\bibliography{IEEEabrv,info21}

\begin{thebibliography}{10}
\providecommand{\url}[1]{#1}
\csname url@samestyle\endcsname
\providecommand{\newblock}{\relax}
\providecommand{\bibinfo}[2]{#2}
\providecommand{\BIBentrySTDinterwordspacing}{\spaceskip=0pt\relax}
\providecommand{\BIBentryALTinterwordstretchfactor}{4}
\providecommand{\BIBentryALTinterwordspacing}{\spaceskip=\fontdimen2\font plus
\BIBentryALTinterwordstretchfactor\fontdimen3\font minus
  \fontdimen4\font\relax}
\providecommand{\BIBforeignlanguage}[2]{{%
\expandafter\ifx\csname l@#1\endcsname\relax
\typeout{** WARNING: IEEEtran.bst: No hyphenation pattern has been}%
\typeout{** loaded for the language `#1'. Using the pattern for}%
\typeout{** the default language instead.}%
\else
\language=\csname l@#1\endcsname
\fi
#2}}
\providecommand{\BIBdecl}{\relax}
\BIBdecl

\bibitem{potaroo}
G.~Huston, ``\url{http://bgp.potaroo.net}.''

\bibitem{filsfils2011bgp}
C.~Filsfils, P.~Mohapatra, J.~Bettink, P.~Dharwadkar, P.~De~Vriendt, Y.~Tsier,
  V.~Van Den~Schrieck, O.~Bonaventure, and P.~Francois, ``Bgp prefix
  independent convergence (pic) technical report,'' \emph{Cisco, Tech. Rep,
  Tech. Rep}, 2011.

\bibitem{Agarwal04controllinghot}
S.~Agarwal, A.~Nucci, and S.~Bhattacharyya, ``Controlling hot potatoes in
  intradomain traffic engineering,'' SPRINT ATL res. rep. RR04-ATL-070677,
  Tech. Rep., 2004.

\bibitem{networkfailures}
A.~{Markopoulou}, G.~{Iannaccone}, S.~{Bhattacharyya}, C.~{Chuah},
  Y.~{Ganjali}, and C.~{Diot}, ``{Characterization of Failures in an
  Operational IP Backbone Network},'' \emph{IEEE/ACM Transactions on
  Networking}, vol.~16, no.~4, pp. 749--762, Aug 2008.

\bibitem{merindol_fine-grained_2018}
P.~Merindol, P.~David, J.-J. Pansiot, F.~Clad, and S.~Vissicchio, ``{A
  Fine-Grained Multi-source Measurement Platform Correlating Routing
  Transitions with Packet Losses},'' \emph{Computer Communications}, vol. 129,
  pp. 166 -- 183, 2018.

\bibitem{raj_survey_2007}
A.~Raj and O.~C. Ibe, ``{A Survey of IP and Multiprotocol Label Switching Fast
  Reroute Schemes},'' \emph{Computer Networks}, vol.~51, no.~8, pp. 1882 --
  1907, 2007.

\bibitem{franccois2014topology}
A.~Bashandy, C.~Filsfils, B.~Decraene, S.~Litkowski, P.~Francois,
  daniel.voyer@bell.ca, F.~Clad, and P.~Camarillo, ``{Topology Independent Fast
  Reroute using Segment Routing},'' Working Draft, IETF Secretariat,
  Internet-Draft draft-bashandy-rtgwg-segment-routing-ti-lfa-05, October 2018.

\bibitem{cardona_bringing_2015}
J.~C. Cardona, P.~Francois, B.~Decraene, J.~Scudder, A.~Simpson, and K.~Patel,
  ``{Bringing High Availability to BGP},'' \emph{Comput. Netw.}, vol.~91,
  no.~C, p. 788–803, Nov. 2015.

\bibitem{teixeira_impact_2008}
R.~Teixeira, A.~Shaikh, T.~Griffin, and J.~Rexford,
  ``\BIBforeignlanguage{en}{Impact of {Hot}-{Potato} {Routing} {Changes} in
  {IP} {Networks}},'' \emph{\BIBforeignlanguage{en}{IEEE/ACM Transactions on
  Networking}}, vol.~16, no.~6, pp. 1295--1307, Dec. 2008.

\bibitem{eventimpact}
F.~Wang, Z.~M. Mao, J.~Wang, L.~Gao, and R.~Bush, ``{A Measurement Study on the
  Impact of Routing Events on End-to-End Internet Path Performance},'' in
  \emph{Proceedings of the 2006 Conference on Applications, Technologies,
  Architectures, and Protocols for Computer Communications}, ser. SIGCOMM
  '06.\hskip 1em plus 0.5em minus 0.4em\relax New York, NY, USA: Association
  for Computing Machinery, 2006, pp. 375--386.

\bibitem{chuah_measuring_nodate-1}
C.-N. Chuah, S.~Bhattacharyya, and C.~Diot, ``\BIBforeignlanguage{en}{Measuring
  {I}-{BGP} {Updates} and {Their} {Impact} on {Traffic}},'' SPRINT ATL res.
  rep. TR02-ATL-051099, Tech. Rep., 1999.

\bibitem{pradeep2012reducing}
K.~Pradeep, O.~Alani \emph{et~al.}, ``{Reducing BGP Convergence Time by Fine
  Tuning the MRAI Timer on Different Topologies},'' in \emph{13th Annual Post
  Graduate Symposium on the Convergence of Telecommunications, Networking and
  Broadcasting PGNET 2012}.\hskip 1em plus 0.5em minus 0.4em\relax The School
  of Computing and Mathematical Sciences, Liverpool John Moores, 2012.

\bibitem{maccari2016pop}
L.~Maccari and R.~L. Cigno, ``{Pop-routing: Centrality-based Tuning of Control
  Messages for Faster Route Convergence},'' in \emph{IEEE INFOCOM 2016-The 35th
  Annual IEEE International Conference on Computer Communications}.\hskip 1em
  plus 0.5em minus 0.4em\relax IEEE, 2016, pp. 1--9.

\bibitem{bremler2003improved}
A.~Bremler-Barr, Y.~Afek, and S.~Schwarz, ``{Improved BGP Convergence via Ghost
  Flushing},'' in \emph{IEEE INFOCOM 2003. Twenty-second Annual Joint
  Conference of the IEEE Computer and Communications Societies (IEEE Cat. No.
  03CH37428)}, vol.~2.\hskip 1em plus 0.5em minus 0.4em\relax IEEE, 2003, pp.
  927--937.

\bibitem{pei2005bgp}
D.~Pei, M.~Azuma, D.~Massey, and L.~Zhang, ``{BGP-RCN: Improving BGP
  Convergence through Root Cause Notification},'' \emph{Computer Networks},
  vol.~48, no.~2, pp. 175--194, 2005.

\bibitem{chandrashekar2005limiting}
J.~Chandrashekar, Z.~Duan, Z.-L. Zhang, and J.~Krasky, ``Limiting path
  exploration in bgp,'' in \emph{Proceedings IEEE 24th Annual Joint Conference
  of the IEEE Computer and Communications Societies.}, vol.~4.\hskip 1em plus
  0.5em minus 0.4em\relax IEEE, 2005, pp. 2337--2348.

\bibitem{holterbach2019blink}
T.~Holterbach, E.~C. Molero, M.~Apostolaki, A.~Dainotti, S.~Vissicchio, and
  L.~Vanbever, ``{Blink: Fast Connectivity Recovery Entirely in the Data
  Plane},'' in \emph{16th {USENIX} Symposium on Networked Systems Design and
  Implementation ({NSDI} 19)}.\hskip 1em plus 0.5em minus 0.4em\relax Boston,
  MA: USENIX Association, Feb. 2019, pp. 161--176.

\bibitem{holterbach2017swift}
T.~Holterbach, S.~Vissicchio, A.~Dainotti, and L.~Vanbever, ``{SWIFT:
  Predictive Fast Reroute},'' in \emph{Proceedings of the Conference of the ACM
  Special Interest Group on Data Communication}, ser. SIGCOMM ’17.\hskip 1em
  plus 0.5em minus 0.4em\relax New York, NY, USA: Association for Computing
  Machinery, 2017, p. 460–473.

\bibitem{kushman2007r}
N.~Kushman, S.~Kandula, and B.~M. Maggs, ``{R-BGP: Staying Connected in a
  Connected World},'' in \emph{4th {USENIX} Symposium on Networked Systems
  Design \& Implementation ({NSDI} 07)}.\hskip 1em plus 0.5em minus 0.4em\relax
  Cambridge, MA: USENIX Association, Apr. 2007.

\bibitem{pelsser_improving_2008}
C.~Pelsser, T.~Takeda, E.~Oki, and K.~Shiomoto,
  ``\BIBforeignlanguage{en}{Improving {Route} {Diversity} through the {Design}
  of {iBGP} {Topologies}},'' in \emph{\BIBforeignlanguage{en}{2008 {IEEE}
  {International} {Conference} on {Communications}}}.\hskip 1em plus 0.5em
  minus 0.4em\relax Beijing, China: IEEE, 2008, pp. 5732--5738.

\bibitem{gamperli2014evaluating}
A.~G{\"a}mperli, V.~Kotronis, and X.~Dimitropoulos, ``{Evaluating the Effect of
  Centralization on Routing Convergence on a Hybrid BGP-SDN Emulation
  Framework},'' \emph{ACM SIGCOMM Computer Communication Review}, vol.~44,
  no.~4, pp. 369--370, 2014.

\bibitem{caesar2005design}
M.~Caesar, D.~Caldwell, N.~Feamster, J.~Rexford, A.~Shaikh, and J.~van~der
  Merwe, ``{Design and Implementation of a Routing Control Platform},'' in
  \emph{Proceedings of the 2nd Conference on Symposium on Networked Systems
  Design \& Implementation - Volume 2}, ser. NSDI’05.\hskip 1em plus 0.5em
  minus 0.4em\relax USA: USENIX Association, 2005, p. 15–28.

\bibitem{simpson2014best}
J.~Uttaro, P.~Francois, K.~Patel, J.~Haas, A.~Simpson, and R.~Fragassi, ``{Best
  Practices for Advertisement of Multiple Paths in IBGP},'' Working Draft, IETF
  Secretariat, Internet-Draft draft-ietf-idr-add-paths-guidelines-08, April
  2016.

\bibitem{vissicchio2014ibgp}
S.~Vissicchio, L.~Cittadini, and G.~Di~Battista, ``{On IBGP Routing
  Policies},'' \emph{IEEE/ACM Trans. Netw.}, vol.~23, no.~1, p. 227–240, Feb.
  2015.

\bibitem{disrupt}
R.~Teixeira and J.~Rexford, ``{Managing Routing Disruptions in Internet Service
  Provider Networks},'' \emph{Comm. Mag.}, vol.~44, no.~3, pp. 160--165, Mar.
  2006.

\bibitem{sobrinho_distributed_2014}
J.~a.~L. Sobrinho, L.~Vanbever, F.~Le, and J.~Rexford, ``{Distributed Route
  Aggregation on the Global Network},'' in \emph{Proceedings of the 10th ACM
  International on Conference on Emerging Networking Experiments and
  Technologies}, ser. CoNEXT ’14.\hskip 1em plus 0.5em minus 0.4em\relax New
  York, NY, USA: Association for Computing Machinery, 2014, p. 161–172.

\bibitem{thorup2001compact}
M.~Thorup and U.~Zwick, ``{Compact Routing Schemes},'' in \emph{Proceedings of
  the Thirteenth Annual ACM Symposium on Parallel Algorithms and
  Architectures}, ser. SPAA ’01.\hskip 1em plus 0.5em minus 0.4em\relax New
  York, NY, USA: Association for Computing Machinery, 2001, p. 1–10.

\bibitem{masterthesis}
\BIBentryALTinterwordspacing
J.-R. Luttringer and P.~Mérindol, ``{OPTIC: An Efficient BGP Protection
  Technique For Optimal Intra-domain Convergence},'' Aug. 2019. [Online].
  Available: \url{https://doi.org/10.5281/zenodo.4436109}
\BIBentrySTDinterwordspacing

\bibitem{luckie_as_2013}
M.~Luckie, B.~Huffaker, A.~Dhamdhere, V.~Giotsas, and k.~claffy,
  ``\BIBforeignlanguage{en}{{AS Relationships, Customer Cones, and
  Validation}},'' in \emph{\BIBforeignlanguage{en}{Proceedings of the 2013
  conference on {Internet} measurement conference - {IMC} '13}}.\hskip 1em plus
  0.5em minus 0.4em\relax Barcelona, Spain: ACM Press, 2013, pp. 243--256.

\bibitem{zenodo}
\BIBentryALTinterwordspacing
Q.~Bramas, P.~Mérindol, C.~Pelsser, and J.-R. Luttringer, ``{A
  Fast-Convergence Routing of the Hot-Potato: The Tool to Perform your own
  Evaluation},'' Feb. 2020. [Online]. Available:
  \url{https://doi.org/10.5281/zenodo.3972128}
\BIBentrySTDinterwordspacing

\end{thebibliography}




\end{document}